\documentclass[authoryear, 5p]{elsarticle}
\usepackage[colorlinks]{hyperref}
\usepackage[latin1]{inputenc}       
\usepackage[T1]{fontenc}            
\usepackage{graphicx}        
\usepackage{multicol}               
\usepackage{multirow}
\usepackage{booktabs}               
\usepackage{appendix}

\begin{document}

\begin{frontmatter}

\title{\small This paper has been published in Ecological Complexity, 2010, 7, 91-99, \href{http://dx.doi.org/10.1016/j.ecocom.2009.07.001}{doi:10.1016/j.ecocom.2009.07.001}\\[0cm] \LARGE Stay by thy neighbor? Social organization determines the efficiency of biodiversity markets with spatial incentives}

\journal{Ecological Complexity}
\author[UFZ]{Florian Hartig\corref{cor}}
\cortext[cor]{Corresponding author, Tel: +49-341-235-1716, Fax: +49-341-235-1473, \href{http://www.ufz.de/index.php?de=10623}{http://www.ufz.de/index.php?de=10623}}
\ead{florian.hartig@ufz.de}
\author[UFZ]{Martin Drechsler}
\ead{martin.drechsler@ufz.de}

\address[UFZ]{UFZ - Helmholtz Centre for Environmental Research, Department of Ecological Modelling, Permoserstr. 15, 04318 Leipzig, Germany}

\begin{abstract}
Market-based conservation instruments, such as payments, auctions or tradable permits, are environmental policies that create financial incentives for landowners to engage in voluntary conservation on their land. But what if ecological processes operate across property boundaries and land use decisions on one property influence ecosystem functions on neighboring sites? This paper examines how to account for such spatial externalities when designing market-based conservation instruments. We use an agent-based model to analyze different spatial metrics and their implications on land use decisions in a dynamic cost environment. The model contains a number of alternative submodels which differ in incentive design and social interactions of agents, the latter including coordinating as well as cooperating behavior of agents. We find that incentive design and social interactions have a strong influence on the spatial allocation and the costs of the conservation market.
\end{abstract}

\begin{keyword}
land use, spatial externalities, spatial incentives, market-based instruments, biodiversity conservation, agent-based models 
\PACS 87.23.Ge, 89.60.Fe, 89.65.-s, 89.65.Gh  \
\end{keyword}
\end{frontmatter}


\section{Introduction}
What is the value of nature? Markets for biodiversity conservation are based on the possibility of rating conservation services (e.g. the provision of an acre of rainforest) in terms of their contribution to conservation goals. A problem that arises when deciding on rating systems is that typical ecological processes operate on a much larger scale than that of typical landowner properties. Therefore, local land use decisions are likely to affect the ecological value of neighboring land. This paper deals with the problem of incorporating such spatial interactions into market-based conservation instruments.\\\\
Market-based instruments have become increasingly popular in recent years \citep{Jack-Designingpaymentsecosystem-2008}, however, they are still a relative new tool for conservation policy. Traditionally, conservation was dominated by regulation and planning approaches which emerged as a response to problems associated with the change and intensification of land use during the last century. Of particular importance for the effectiveness of top-down approaches such as regulations and planning is the inclusion of both the monetary costs and the ecological benefits of conservation measures \citep{Faith-Integratingconservationand-1996, Ando-SpeciesDistributionsLand-1998, Margules-Systematicconservationplanning-2000}. This insight points to a practical problem of planning approaches: local costs are difficult or expensive to estimate,  and it is seldom in the interest of landowners to report them honestly. Moreover, costs may change over time. In these cases, market instruments provide an alternative to planning approaches because they are able to efficiently allocate conservation efforts to the spatial distribution of conservation costs, even when cost information is only available to landowners and not to the regulating authorities (information asymmetry).\\\\
The common principle of market-based instruments is to introduce a metric that rates the value of conservation measures in terms of their contribution to conservation goals. This metric translates conservation measures into one currency (commodification) and thereby makes conservation comparable and tradable based on this currency \citep{Salzman-Currenciesandcommodification-2000, Salzman-Creatingmarketsecosystem-2005}. In practice, different names are used for this currency. We will use the term "credits" throughout this paper, and say that the metric measures the amount of conservation provided by a site in credits. Demand for credits may be created by different mechanisms, e.g. payments \citep{WUNDER-EfficiencyofPayments-2007, Drechsler-model-basedapproachdesigning-2007, Engel-Designingpaymentsenvironmental-2008}, auctions \citep{Latacz-Lohmann-Auctionsasmeans-1998} or biodiversity offset schemes \citep{Panayotou-Conservationofbiodiversity-1994, Chomitz-Transferabledevelopmentrights-2004}. This demand for credits, together with the metric to measure them, creates an incentive for conservation. In a sense, we may view the process of trading credits as a policy-based site selection algorithm \citep{Faith-Complementaritybiodiversityviability-2003}: competition among suppliers automatically extracts the sites that can provide conservation measures at the lowest costs.\\\\ 
Yet, while markets may help to solve the problem of cost information asymmetry between landowners and regulators, the definition of an accurate measure of conservational value runs into problems when the ecological values of sites are dependent on each other. An apparent solution is to incorporate spatial dependencies into market values \citep{Parkhurst-Agglomerationbonusincentive-2002}. While this is generally possible, it implies that conservation decisions may change the market value of neighboring land. In the presence of such spatial interactions, referred to  in the literature as externalities, spillovers or site synergies, markets may fail to create an efficient spatial allocation of the traded good  \citep{Mills-Transferabledevelopmentrights-1980}. Moreover, ecological processes may operate on a large range of spatial and temporal scales and show complex dependencies, making the exact accounting for ecological interactions potentially very difficult. \citet{Moilanen-ReserveSelectionUsing-2005} discusses a case of interactions where already static optimization is computationally hard. In such a case, it is unlikely that market participants would find the optimal allocation of conservation measures, particularly if they are subject to external drivers such as changing conservation costs.\\\\ 
The aim of this paper is to analyze the functioning of simple spatial incentives in market-based conservation instruments. We use an agent-based model to examine whether simple spatial connectivity incentives operating on a local scale can effectively influence the larger scale allocation of conservation measures, and how design of the spatial incentives and social organization affect the emerging landscape structure. 

\section{Problem definition and modeling approach} \label{sec: Spatial Ecology and Spatial Economics}
The fact that sites may interact and influence the ecological value of neighboring sites creates a number of issues which make spatial incentives an interesting problem for economics and conservation research. A number of important real-world processes create spatial interactions between sites. One example is habitat fragmentation, which constitutes a major problem for biodiversity conservation \citep{Saunders-BiologicalConsequencesOf-1991, MA-EcosystemsandHuman-2005}. The origin of this problem is that many species require to travel between habitat patches in the landscape. When habitats are increasingly isolated, e.g. through land use change, they may eventually be of very low value for biodiversity because species can not reach them. Therefore, the ecological value of a natural habitat generally increases when other natural habitats are in the vicinity.\\\\
In this paper, we assume that there is a symmetric, positive interaction benefit between conserved sites. As discussed above, this is very likely to be the case in real-world conservation problems. However, other cases such as non-symmetric benefits \citep[see][]{Vuilleumier-Doescolonizationasymmetry-2006} or negative interactions could equally be targeted with markets. The aim of this section is to clarify the conceptual questions that arise from including spatial incentives in market instruments, and to formulate more precisely the questions we want to answer with the model. 
\subsection{Marginal and additive incentives}
\begin{figure}[]
\begin{center}
\includegraphics[width=6cm]{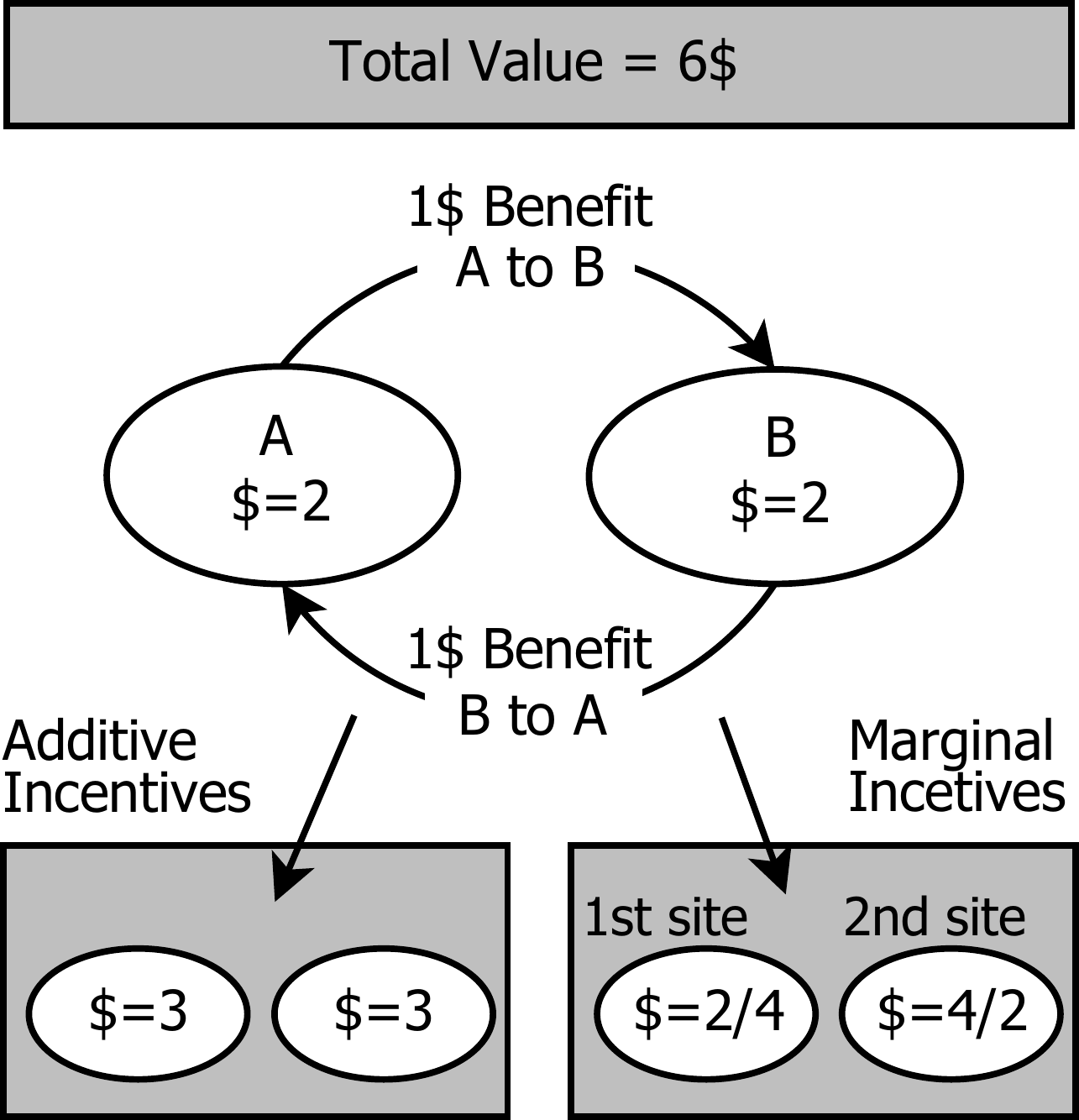}
\caption{Two sites of equal size, A and B, are of mutual benefit to each other. When isolated, each site has an ecological value of 2 each. Connecting the two sites creates an additional value of 2, the results of a benefit of 1 in each direction. Given additive incentives, each landowner receives 3 credits, while under marginal incentives, the first one to come would receive 2 credits, whilst the second receives 4 credits because his site adds all the mutual benefits and thus confers a marginal value of 4 to the network. Yet, note that if both sites are present, each site has a  marginal value of 4. Therefore, the first site to move would create a marginal damage of 4, while the damage of removing the second site would only be 2, as indicated by the second number.} \label{figure: neighborbenefits}
\end{center}
\end{figure}

The first question relates to the difference between the total and the local valuation of conservation measures. Let us assume that we have a market instrument with spatial incentives, and we have a metric $U$ which measures the ecological value of a landscape and includes spatial interactions between sites. As an example, we may have two sites which have, as isolated sites, an ecological value of $2$ credits each. If each site benefits from the presence of the other, the total ecological value of the two sites will be higher than the sum of the single sites, which is $2+2 = 4$ credits. Let us assume that the collective value $U$ is $6$ credits, consisting of the single values which amounted to $4$ credits, and $2$ additional credits originating from the positive interactions. What is the value of these connected sites? One may assume that, as both sites are identical, it should be $3$ for each site. However, removing any of the two connected sites would leave us with a single site of value $2$, suggesting that the value of the first site which was removed was in fact $4$. An illustration of this is given in Fig.~\ref{figure: neighborbenefits}.\\\\
To give a more mathematical description of this, assume that the metric measuring the credits awarded to a site as follows
\begin{equation}\label{eq. global value general}
    U = \sum_i \left\{(1-m)\cdot A_i  +\;  m \cdot \beta_i\right\}.
\end{equation} 
Here, the sum runs over all conserved sites, and each site is evaluated according to its area $A_i$ and its connectivity $\beta_i$, i.e. its connection and therefore its interactions with neighboring sites. For now, we view $\beta_i$ as a generic measure of connectivity, but we will give a specific expression in the model description. The parameter $m$, taking values from 0 to 1, specifies the trade-off between area $A_i$ and connectivity $\beta_i$. We call $m$ the connectivity weight. If $m=0$, a connected site receives the same amount of credits as an unconnected site. Increasing $m$, connectivity becomes more important. If $m=1$, only the connectivity of sites is rewarded with credits. Despite its simplicity, this function essentially captures the trade-off between area and connectivity, which is found to be very important for the conservational value of sites in fragmented landscapes \citep[e.g.][]{Frank-formulameanlifetime-2002, Drechsler-Predictingmetapopulationlifetime-2009}. The appropriate value of $m$ depends on the species under consideration. For some systems, connectivity is more important, while others depend mostly on area.\\\\ 
\textbf{Marginal incentives}~(1): The change of total ecological value originating from the removal of one site is called the marginal value $b^{mar}$ of this site. For our former example, each of the two connected sites has a marginal value of $4$, because their removal would decrease the total value $U$ by $4$. Assuming eq.~\ref{eq. global value general}, the marginal value of the i-th site is given by 
\begin{equation}\label{eq: marginal Ecological Benefit}
    b^{mar}_i = (1-m)\cdot A_i + 2 \cdot m \cdot \beta_i \;.
\end{equation}
The factor $2$ in the second term originates from the fact that not only the connectivity of the focal site is removed, but also the connectivity of the sites it interacts with is decreased by the same value (symmetric benefits). Incentives based on the marginal value we call "marginal incentives". There are two important points to note about marginal incentives. Firstly, if more than one site is changed, the marginal values of the sites depend on the order of trading. Assuming a positive interaction between sites, there is a first-mover disadvantage: the marginal value of the first site created is less than the marginal value of the second site, and the costs for the first site to be removed are higher than for the second site (Fig.~\ref{figure: neighborbenefits} ). Secondly, the marginal value of two sites is generally not the sum of the two sites' marginal values. Marginal values do not add up to the total value. These two properties have strong implications for conservation policy. For using marginal incentives, we have to make sure that the order of trading is known, i.e. that trading takes place sequentially. Moreover, another problem arises when landowners want to withdraw a site from the market, and this sites marginal value has changed since creation because other sites have been added in the vicinity. Marginal incentives require that the costs for withdrawing such a site must be higher than the benefits initially awarded for it (see Fig.~\ref{figure: neighborbenefits} ). This means that marginal incentives are difficult to use for short term incentive mechanisms such as yearly flat rate payments. They are, however, very suitable for markets such as tradable permits where both creation and destruction is targeted, or markets with long-term contracts, where the problem of habitat destruction does not arise.\\\\
\textbf{Additive incentives}~(2): Another way to arrive at a sensible local evaluation of sites is by simply dividing mutual benefits equally among the involved sites (see Fig.~\ref{figure: neighborbenefits}). The value that results from sharing the benefits is still higher than the value of isolated sites, but it is generally lower than the marginal value. For the case of eq.~\ref{eq. global value general}, equally sharing the mutual benefits results in local values of   
\begin{equation}\label{eq: additive eco benefit}
    b^{add}_i \equiv (1-m) \cdot A_i + m \cdot \beta_i \;.
\end{equation}
We call these incentives additive incentives because their sum adds up to the total value: 
\begin{equation}\label{eq:sum local Benefit constraint}
    \sum_i b_i^{add} = U \;.
\end{equation}
Markets with additive spatial incentives have been suggested, e.g. by \citet{Parkhurst-Agglomerationbonusincentive-2002} or \citet{Hartig-Smartspatialincentives-2009}, as a means to improve the spatial agglomeration of conservation sites. Additive incentives have the advantage that the order of creation or destruction has no impact on the value of a particular site, and the value of any number of sites is simply the sum of the values of these sites. One drawback, however, is that a change of conservation on one site may directly affect the ecological value and therefore the credits awarded to neighboring conserved sites. This might create acceptance problems for real-world conservation schemes. Another issue is that marginal values assign only a part of the costs and benefits created by a land use change to its originator. This may lead to efficiency losses. We show evidence for this in the results.

\subsection{Social Organization}\label{sec: bevaioral assumptions}
If market incentives for conserving a site depend on the neighboring sites, it is important for landowners to know their neighbors' intentions. We assume that agents are always profit maximizing and myopic in the sense that they base their decision on the most profitable action for the next timestep without displaying strategic behavior. Within this setting, we consider three behavioral submodels:\\\\
\textbf{Null model}~(a): In the null model, agents observe the present landscape configuration and decide based on the prospect of the future land configuration being the same as the present one.\\\\
\textbf{Coordination through cheap talk}~(b): We assume that agents may communicate their future intentions. This can be beneficial because it increases the accuracy of the estimate about the upcoming land configuration. We understand coordination as the communication of non-binding information about the present state of the decision, often called cheap talk \citep{Farrell-Cheaptalk-1996}. Experimental studies have shown that the possibility of coordination by cheap talk leads to an increased probability of finding cost-effective configurations \citep{Parkhurst-Agglomerationbonusincentive-2002}. Generally, cheap talk also includes the possibility of strategic lies. This option is omitted in the simulation and hence agents will always stick to the action they communicated as long as their information does not change.\\\\
\textbf{Cooperation}~(c): Further payoff improvements are possible if agents not only coordinate, but cooperate. Cooperation means that conservation is provided if it is beneficial for the group, even if this does not maximize the profits of each individual. As marginal and additive incentives differ only in how benefits are distributed to individuals, but not in how many credits are rewarded in total, there is no difference between them from the point of view of cooperating agents. In a cooperating group, agents reveal their true costs, the group chooses the best configuration of sites, and the payoffs are distributed among the group members according to their costs. It would be possible that single individuals exploit such a system by communicating higher than the true costs to increase their share from the group benefits. We do not consider this possibility in the model, but we will address this issue in the discussion. \\\\
For options a) and b), decisions are made individually, while cooperating agents c) decide collectively. We will later see that these structurally different decision processes are also reflected in the resulting land use pattern. Table~\ref{table: overview Game types} summarizes the possible combinations of submodels and their properties.

\begin{table*}[]
  \centering
  \begin{tabular}{lllll}\toprule
 \textsc{Label }& \textsc{Submodel }& \textsc{Decisions }   &   Time order &\textsc{Communication } \\\midrule \addlinespace[0.2cm] 
 (a.1) &  Null marginal &       \multirow{4}{*}{individually} &  dependent &  \multirow{2}{*}{none}\\
 (a.2) &  Null additive &                                   &  independent &   \\
 (b.1) &  Coordination marginal&                            &  dependent& \multirow{2}{*}{intentions} \\ 
 (b.2) &  Coordination additive&                            &  independent&   \\ \midrule 
  (c)  &  Cooperation & Collectively                          & independent  &  full information \\ \bottomrule

\end{tabular}
  \caption{Overview of alternative submodels for the spatial incentives and for agent behavior}\label{table: overview Game types}
\end{table*}

\section{Model description} \label{sec: model description}
\subsection{Overview and purpose} 
The aim of the model is to examine the effect of spatial incentives and social organization on the emerging landscape structures and on the total cost of a conservation market. The model is based on the models introduced in \citet{Drechsler-Applyingtradablepermits-2009} and \citet{Hartig-Smartspatialincentives-2009}. It predicts the spatial allocation of conservation sites emerging from agents' decisions to conserve their land or not. These decisions are driven by spatially and temporally heterogeneous costs of conservation and by the benefits of conservation, which depend on the current market price and on the amount of conserved cells in the neighborhood. This dependence on the state of the neighboring cells resembles 2-dimensional spin models with local interactions which have been used to analyze phenomena of social interactions \citep{Galam-Fromindividualchoice-2000, Sznajd-Weron-Opinionevolutionin-2000, Holyst-Phasetransitionsin-2000, Schweitzer-Coordinationofdecisions-2002}. It also exhibits similarities to the Random Field Ising Model \citep{Imry-Random-FieldInstabilityof-1975} and non-equilibrium models such as \citet{Hausmann-Stationarypropertiesof-1997} and \citet{Acharyya-Nonequilibriumphasetransition-1998}. In the follwowing subsections, we give an overview first of state variables and scales, and then about the processes implemented in the model.   

\subsection{State variables and scales} 
The simulation is carried out on a 2-dimensional grid with 50x50 grid cells (sites) and periodic boundary conditions. Every grid cell $x_i$ is owned by a different landowner (agent) and can be occupied with a habitat ($\sigma_i = 1$) or be used for other purposes ($\sigma_i = 0$). Although the model may be applied to any spatial and temporal scale, we think of grid cells as being of the size of an average agricultural field in Europe, and time steps being a year. The occupancy of a grid cell results in conservation costs $c_i(t)$ which may be different for each cell.\\\\
We assume that we have a market instrument which rewards a certain price $P$ for each conservation credit produced by landowners. Conservation credits are calculated according to the metric given in eq.~\ref{eq. global value general}, where the area $A_i$ was set to unity and is therefore omitted.
\begin{equation}\label{eq. global value specific}
    U = \sum_i^N \sigma_i \left\{(1-m)  +\;  m \cdot \beta_i\right\}
\end{equation} 
As before, $m$ weights the importance of connectivity $ \beta_i $ relative to area for the ecosystem function which is targeted by the market. The connectivity metric $ \beta_i $, which measures the interactions between patches, is chosen as 
\begin{equation}\label{eq: connectivity metric}
     \beta_i \equiv \frac{1}{8}\sum_{<j>_i} \sigma_j
\end{equation}
where $<j>_i$ indicates all cells $j$ which belong to the 8 cells $x_j$ in a Moore neighborhood of $x_i$. The metric is normalized to one and basically measures the fraction of the 8 cells in the neighborhood of the focal cell which are used for conservation. We assume a totally inelastic demand 
\begin{equation}\label{eq: Target Value}
    D = \lambda \cdot N 
\end{equation}
for conservation credits, which should equal the supply $U$. A list of all basic state variables and parameters is given in Table~ \ref{table:state variables}. We define the average number of occupied sites
\begin{equation}\label{eq: def alpha and beta}
   \alpha \equiv \frac{1}{N} \sum_{i=1}^N \sigma_i \;.
\end{equation}
Further, we define the average connectivity $K$ of the occupied sites as
\begin{equation}\label{def:connectivity}
   K \equiv \frac{1}{\alpha \cdot N}\sum_{i=1}^N \sigma_i\cdot \beta_i
\end{equation}
and total conservation costs $C$ of a landscape configuration as the sum of the costs of all conserved grid cells, divided by the number of grid cells
\begin{equation}\label{def:societal costs}
   C \equiv \frac{1}{N} \sum_{i=1}^N \sigma_i \cdot c_i \;.
\end{equation}

\begin{table}[b]
  \centering
  \begin{tabular}{lll} \toprule
  \textsc{Symbol} & \textsc{Connotation} & \textsc{Range}  \\ \midrule \addlinespace[0.2cm] 

  $x_i$ & i-th cell on the grid &  \\ 
  $\sigma_i$ & State of the i-th cell  &  $\{0,1\}$ \\
  $c_i(t)$ & Costs of $\sigma_i$ = 1 at t&  $[1-\delta ... 1+\delta]$  \\ 
  $P$ & Market price & $[1-\delta ... 1+\delta]$ \\ 
  \multicolumn{3}{l} { }  \\
  $\delta$ & Cost heterogeneity & $[0..1]$ \\
  $m$ & Connectivity weight  & $[0..1]$ \\  
  $\lambda $ & Fixed demand & $[0..1]$\\ \bottomrule

\end{tabular}
  \caption{List of state variables (top) and parameters (bottom).}\label{table:state variables}
\end{table}
\subsection{Process overview and scheduling:}
\begin{figure}[]
\begin{center}
\includegraphics[width=8cm]{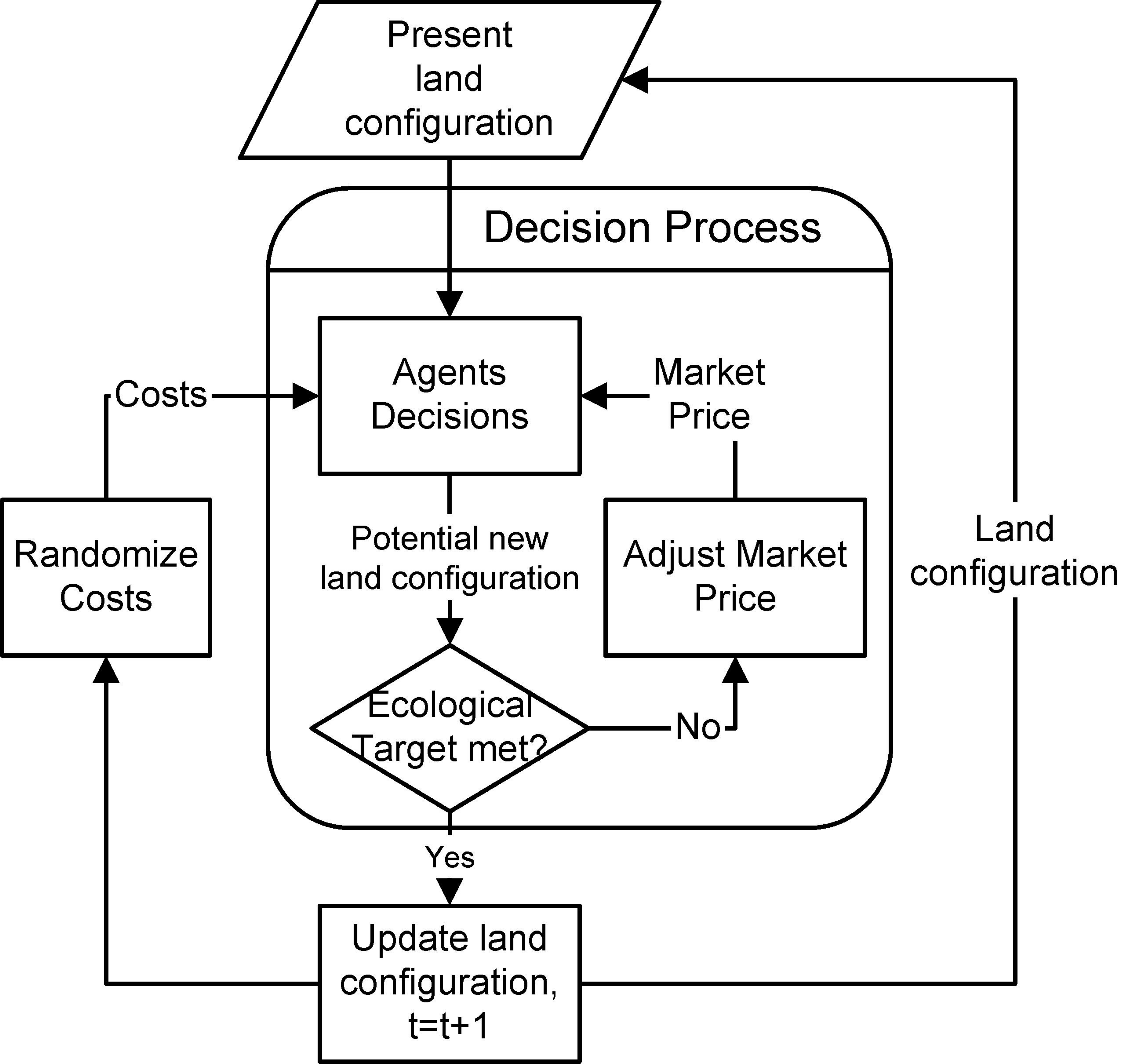}
\caption{Flow diagram of processes within one timestep. \hspace{5cm}} \label{figure:program flow}
\end{center}
\end{figure}

At each time step, costs $c_i$ are drawn from a uniform distribution of mean 1 and width $2\delta$. Agents decide to maintain a site as habitat based on their costs $c_i$, the market price $P$ and the estimated credit value $\tilde{b}_i$. They maintain a habitat on $x_i$ at timestep $t$ if conservation yields a positive net benefit $\pi$
\begin{equation}\label{eq: profit bedingung}
   \pi_i \equiv - c_i(t) + P \cdot \tilde{b}_i(t) > 0.
\end{equation}
The credits rewarded to agents are either the marginal incentives (eq.~\ref{eq: marginal Ecological Benefit}) given by $(1-m) + 2 m \cdot \beta$ or the additive incentives (eq.~\ref{eq: additive eco benefit}) given by $(1-m) + m \cdot \beta$. The final benefits $b_i$ rewarded at the end of the round can differ from the estimated benefits $\tilde{b}_i$ because subsequent decisions by other agents can change the landscape configuration. The accuracy of the estimate $\tilde{b}_i$ depends on the applied behavior submodel. The three submodels discussed in section~\ref{sec: bevaioral assumptions} are implemented in the following way:\\\\
\textbf{Null model}: All agents decide in parallel according to eq.~\ref{eq: profit bedingung} without being informed about the decisions of other agents at this timestep.\\
\textbf{Coordination through cheap talk}: Agents decide sequentially in random order according to eq.~\ref{eq: profit bedingung}. After each decision, all agents are informed about the new configuration. This procedure is repeated a number of times, mimicking the outcome of a non-binding exchange of information.\\
\textbf{Cooperation}: Cooperation is modeled by global optimization with full information about credit benefits and conservation costs. The details of the optimization procedure are described in Appendix~\ref{sec: Optimization}.\\\\
Step by step, the decisions of all agents are collected and the resulting ecological value $U$ as given in eq.~\ref{eq. global value specific} is compared with the demand $D$ (eq.~\ref{eq: Target Value}). The emergence of an equilibrium between demand and supply is modeled by repeatedly adjusting the market price $P$ in eq.~\ref{eq: profit bedingung} until demand and supply are balanced. The order in which agents are asked is randomized at every timestep, but does not change while the market price is adjusted. Fig.~\ref{figure:program flow} shows a flow diagram of the processes within one timestep.\\\\
Simulation runs were initialized with a random landscape configuration which delivered the target supply of conservation. A series of tests showed that the initial configuration has no influence on the resulting landscape after several hundred time steps. This holds true also for non-random start configurations. Data acquisition was started after 600 trading steps to ensure that the simulation had reached its steady state. The results show the mean of 50 runs. Standard deviations for all values were calculated but omitted in the figures because they were very small. 

\section{Analytical approximations -- clusters and disorder}
Although the number of landscape configurations which fulfill the conservation target is extremely large, two structures are particularly simple and their properties may be calculated analytically: one is a landscape where habitats are concentrated into one big cluster, the other is a landscape where habitats are scattered according to the lowest costs. Examples of these states will appear later in the results (Fig.~\ref{figure: uebersicht connectivity mit contour plots}). Clustering and disorder mark the extreme cases of possible connectivity values: No other state produces more connectivity than a cluster, and no reasonable state (leaving aside anticorrelated structures) produces less connectivity than the scattered, disordered state. It will prove useful for the interpretation of the results to derive some analytical approximations for these two states.

\subsection{Critical $\delta$ value for clustering}
Let us assume that agents can only choose between clustering and spread. Clustered structures lead to a higher credit value per cell, but also to higher average costs, because a spread, disordered configuration can more effectively allocate conservation efforts on the sites with the lowest costs. At low values of cost heterogeneity $\delta$ compared to connectivity weight $m$, a clustered structure is clearly favored. At increasing cost heterogeneity, we expect a critical value $\delta^c$ where the net benefits from clustering become smaller than the net benefits from spread. \\\\
We can derive this critical value by equating the worst benefit-cost ratio of cells within a cluster with the best ratio of an isolated cell. In a cluster, the habitats with the highest costs have $c=1+\delta$, while outside the cluster, the cells with lowest costs have $ c=1-\delta$. The ecological value of a clustered and a disordered cell is given by eqs.~\ref{eq: additive eco benefit}~and~\ref{eq: marginal Ecological Benefit}. Hence, we obtain
\begin{equation}\label{eq:theoretical curves}
    \frac{1}{1+\delta^c_{\mathrm{add}}}=\frac{1-m}{1-\delta^c_{\mathrm{add}}} \Rightarrow \delta^c_{\mathrm{add}} = m/(2-m)
\end{equation}
\begin{equation}\label{eq:theoretical curves 2}
    \frac{1+m}{1+\delta^c_{\mathrm{mar}}}=\frac{1-m}{1-\delta^c_{\mathrm{mar}}} \Rightarrow \delta^c_{\mathrm{mar}} = m 
\end{equation}
as critical values for additive and marginal incentives, respectively.
\subsection{Clustered and disordered cost level}
Further, we are also interested in the costs of maintaining the land at either of the two states. Cells in a cluster have 8 neighbors and therefore yield an average ecological value of $(1-m) + m \cdot \beta = 1$ per cell (eq.~\ref{eq. global value specific}). Therefore, a number of $\lambda \cdot N$ patches satisfies the fixed demand of $\lambda$ per cell (eq.~\ref{eq: Target Value}). As costs are spatially uncorrelated, the mean costs $\bar{c}_p$ within a cluster are approximately equal to the mean costs of the landscape (for the chosen function $\bar{c}_p = 1$) as long as finite size effects can be neglected. Thus, the total costs of satisfying the demand of $\lambda \cdot N$ credits through a cluster are 
\begin{equation}\label{eq: Theoretical ordered costs}
    C_{\mathrm{clu}}= \frac{1}{N} \cdot N \cdot \lambda  = \lambda \; .
\end{equation}
This means that the costs for a cluster are simply constant and proportional to its size. In a disordered state of density $\alpha$, occupied sites are distributed randomly according to the lowest costs. Each occupied cell has on average $8 \cdot \alpha$ neighbors, leading to an average credit value of $(1-m)+m \cdot \alpha$ per grid cell. To reach an average credit supply of $\lambda$ per cell, we require that 
\begin{equation}\label{eq: Theoretical alpha disordered}
    \alpha\left[(1-m)+m \cdot \alpha \right]= \lambda \;.
\end{equation}
Solving for $\alpha$ yields
\begin{equation}\label{eq: theoretical disordered level of alpha}
   \alpha = \frac{-1+m+{\sqrt{{{(-1+m)}^2}+4 m\lambda }}}{2 m} \; .
\end{equation}
When the cells with the lowest costs are selected first, the marginal costs $c_p(\zeta)$ increase with the fraction $\zeta$ of cells which are selected. With costs being uniformly distributed across the interval $[1-\delta \dots 1+\delta ]$, $c_p(\zeta)$ is given by
\begin{equation}\label{eq: marginal costs}
    c_p(\zeta)= (1 - \delta) + 2 \delta \cdot \zeta \; .
\end{equation}
From this, we can derive the total costs for a disordered state as
\begin{equation}\label{eq: Theoretical disordered costs}
    C_{\mathrm{do}}= \alpha \int_0^\alpha c_p(\zeta) d\zeta  =  \alpha - (\alpha-\alpha^2) \cdot \delta \; .  
\end{equation}
Thus, the costs for a disordered state are dependent on the cost heterogeneity $\delta$ and are linearly decreasing with $\delta$. The cost lines for the clustered and the disordered state mark an upper boundary for cost-effective configurations in Fig.~\ref{figure: societal costs} and Fig.~\ref{figure: societal costs  coordination}. As a function of $\delta$, they intersect at 
\begin{equation}\label{eq: Theoretical intersection of the cost lines}
    \delta = \frac{\alpha-\lambda}{\alpha-\alpha^2}= m
\end{equation}
where the second equality is derived from inserting the expression in eq.~\ref{eq: theoretical disordered level of alpha} for $\alpha$. Note that the final result, $\delta = m$, coincides with the critical point for marginal incentives (eq.~\ref{eq:theoretical curves 2}). 

\begin{figure*}[t]
\begin{center}
\includegraphics[width= 14cm]{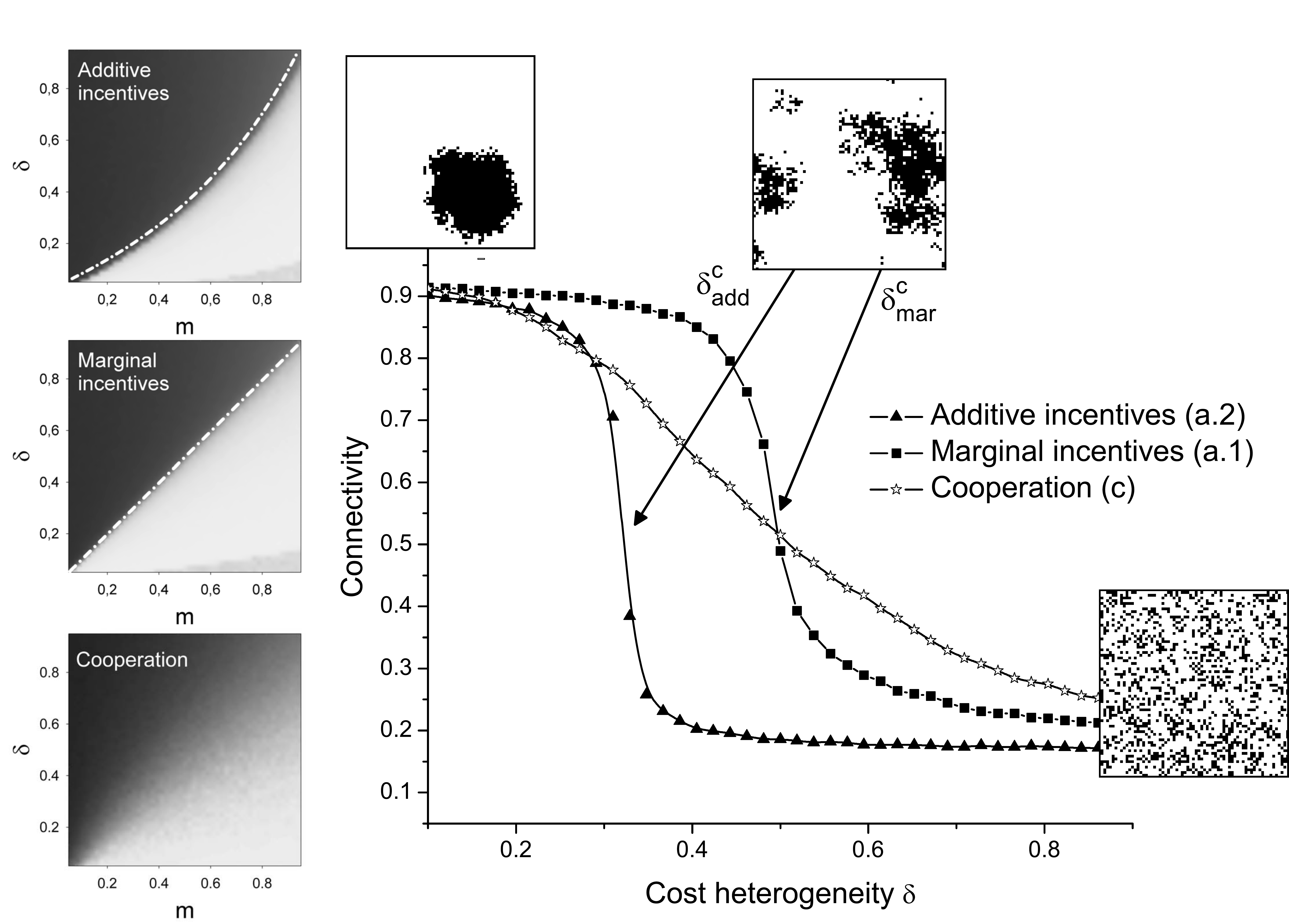}
\caption{On the left, landscape connectivity $K$ (eq.~\ref{def:connectivity})for marginal incentives (top), additive incentives (middle) and cooperation (bottom) as a function of cost heterogeneity $\delta$ and connectivity weight $m$ at $\lambda=0.1$. Darker colors indicate lower connectivity. Dotted lines display the theoretical curves as given in eqs.~\ref{eq:theoretical curves}~and~\ref{eq:theoretical curves 2}. The right graph shows a cross-section of the three plots in $\delta$ direction at $m=0.5$ together with typical landscape structures emerging from the simulations in the three domains: Clusters (left), transition states (middle) and disorder (right). The two arrows indicate the transition state appears at different levels of $\delta$, depending on the chosen incentive mechanism.} \label{figure: uebersicht connectivity mit contour plots}
\end{center}
\end{figure*}

\begin{table}[b]
  \centering
\begin{tabular}{lll}\toprule
  \textsc{Model} & \textsc{Simulation} & \textsc{Approximation} \\ \midrule \addlinespace[0.2cm] 
  Additive Incentives & $0.33\pm 0.02$  & $0.33$  \\ 
  Marginal Incentives & $0.50 \pm 0.02$  & $0.5$  \\ \bottomrule
\end{tabular}
  \caption{Critical value $\delta^c$ (measured as $\delta$ at half transition) at $m=0.5$ from Fig.~\ref{figure: uebersicht connectivity mit contour plots} together with theoretical expectations }\label{table: theoretical vs. simulations values of critical delta}
\end{table}
\section{Simulation results} \label{sec: results}

\subsection{Critical values}
\begin{figure}[]
\begin{center}
\includegraphics[width=8cm]{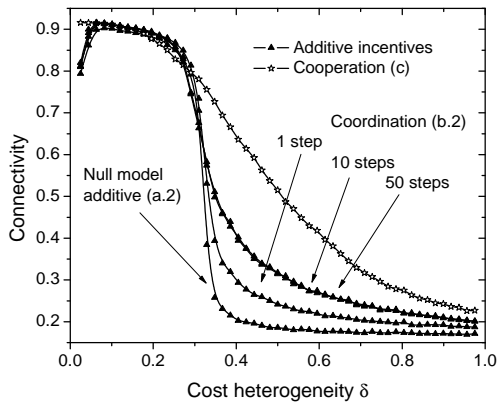}
\caption{Connectivity $K$ (eq.~\ref{def:connectivity}) as a function of cost heterogeneity $\delta$ at a fixed $m = 0.5$. It can be seen that increased communication does not affect the critical point $\delta^c$ of the additive incentive curves.} \label{figure: connectivity coordination}
\end{center}
\end{figure}

When analyzing the parameter space of $m$ and $\delta$, we observe steep transitions of all aggregated state variables along a curve of critical value pairs $(\delta^c, m^c)$. Parameter values beyond this curve lead to disordered landscape structures, while values below $\delta^c$ lead to ordered, connected structures. The shape of the transition curve in the space of $m$ and $\delta$ differs for additive and marginal incentives. Fig.~\ref{figure: uebersicht connectivity mit contour plots} shows the simulation results for additive and marginal incentives at $\lambda = 0.1$ together with the analytically derived curves eqs.~\ref{eq:theoretical curves}~and~\ref{eq:theoretical curves 2}. The  right side of Fig.~\ref{figure: uebersicht connectivity mit contour plots} shows a vertical cross-section of the left hand side plots in $\delta$-direction at $m=0.5$. The transition values $\delta^c$ in this curve agree with the analytical approximations from eqs.~\ref{eq:theoretical curves}~and~\ref{eq:theoretical curves 2} (Table~\ref{table: theoretical vs. simulations values of critical delta}).\\\\
The introduction of coordination (submodel b.1 and b.2) leaves these results largely unchanged. Fig.~\ref{figure: connectivity coordination} shows that coordination only slightly decreases the steepness of the transition, but does not change the transition point. In contrast to that, cooperative decisions lead to a rather broad, nearly linear transition for the same parameter values. 

\subsection{Cost-effectiveness}\label{sec: results / costs}
\begin{figure}[]
\begin{center}
\includegraphics[width=8cm]{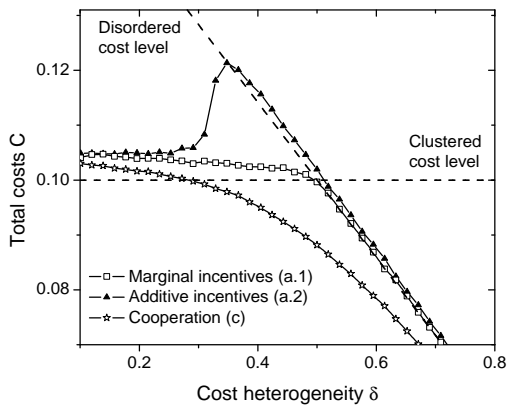}
\caption{Total conservation costs as a function of cost heterogeneity $\delta$ at $m = 0.5$ and $\lambda = 0.1$ (solid lines) together with theoretical costs for the clustered and the disordered state (dashed lines).} \label{figure: societal costs}
\end{center}
\end{figure}

A second question was how the three behavior options would perform in terms of costs needed to reach an equal amount of conservation credits. Fig.~\ref{figure: societal costs} shows the total conservation costs as a function of cost heterogeneity $\delta$ at $m=0.5$, together with the theoretical cost levels for the clustered and the disordered states as calculated in eqs.~\ref{eq: Theoretical ordered costs} and~\ref{eq: Theoretical disordered costs} (dotted lines). Noticeably, the cost function of additive incentives displays a hump around the transition area (Fig.~\ref{figure: societal costs}), leading to approximately $20\%$ higher costs for additive compared to marginal incentives. Considering the theoretical cost functions for the ordered and the disordered states (eqs.~\ref{eq: Theoretical ordered costs},~\ref{eq: Theoretical disordered costs}), the reason becomes evident: Under marginal incentives, agents switch from clustering to disorder right at the point where the costs of a cluster and the costs of a disordered state intersect (eq.~\ref{eq: Theoretical intersection of the cost lines}). In contrast, under additive incentives, agents switch to disordered configurations earlier at a lower level of cost heterogeneity $\delta$. At this point, the cost level of a disordered configuration is still considerably higher than that of a clustered one, resulting in efficiency losses in this area.

\subsection{Influence of behavior on costs}
\begin{figure}[]
\begin{center}
\includegraphics[width=8cm]{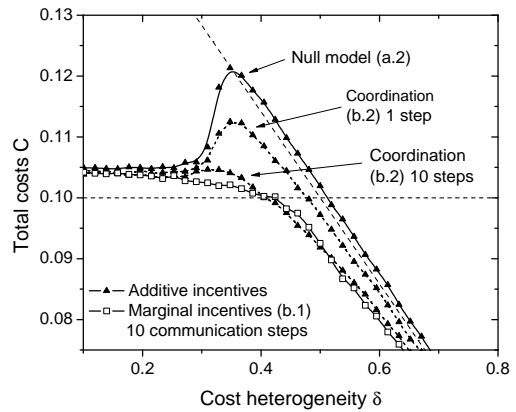}
\caption{Effect of coordination: Plottet are the total conservation costs as a function of cost heterogeneity $\delta$ at $m = 0.5$ and $\lambda = 0.1$. With increasing coordination steps, the costs of additive incentives approach the costs of marginal incentives.} \label{figure: societal costs coordination} 
\end{center}
\end{figure}

The inefficiency which has been observed for additive incentives is to a great extent mitigated by the introduction of coordination (Fig.~\ref{figure: societal costs coordination}). For a value of 10 communication steps per trading period, only small differences between additive and marginal incentives remain. This stems mainly from a better adaptation of land use to the current costs. Transition points and thus the landscape structure remain largely unchanged by coordination as shown in Fig.~\ref{figure:  connectivity coordination}.\\\\
The total conservation costs under cooperation are considerably lower than for the other options (Fig.~\ref{figure: societal costs}). They differ mostly around the critical values $\delta_c$, where optimal decisions are more difficult to obtain than for extreme values of $\delta$ where one allocation pattern (cluster or disorder) is clearly favored. 

\section{Discussion}
\subsection{Main findings}
The aim of this modeling study was to examine the effect of simple spatial incentives and social organization on the emerging landscape structures and on the total cost of a conservation market. We introduced an agent-based model of a conservation market with spatial incentives. The model enabled us to compare different spatial incentives (additive and marginal) and different types of agent behavior, including individual and collective decision processes. By individual decisions, we mean that agents decide individually according to their costs and benefits, while collective decisions consider the costs and benefits of the whole group.\\\\ 
Our results show that rewarding more conservation credits when other conservation sites were directly connected to the focal site could effectively the larger scale structure of conservation measures. The connectivity weight $m$, which determines the amount of credits which are rewarded for conservation, acts as a control parameter that determines the clustering of habitats in the landscape. The exact shape of this response depends on the heterogeneity of conservation costs, the incentive mechanism, and the assumed behavior of agents (Fig.~\ref{figure: uebersicht connectivity mit contour plots}).\\\\
Individual decision processes with low information (null model a.1,a.2) lead to a partitioning of the parameter space, each producing very distinct landscape patterns, separated by a small transition area.  These predominant patterns are a completely clustered state, appearing when cost heterogeneity $\delta$ is low compared to the connectivity weight $m$, and a completely disordered state, emerging when $\delta$ is high compared to $m$. The reason for the stability of these two states in a dynamical and heterogeneous cost background lies in the fact that, starting from one of them, agents repeatedly end up in a very similar configuration, since trading stops as soon as a local cost optimum is found. Decision processes with more information (coordination b.1,b.2) explore a wider range of options, but even in the case of coordination with a large number of communication steps, agents stop trading once trade cannot improve the position of any single agent (Fig.~\ref{figure: connectivity coordination}). Only cooperating agents (submodel c) completely avoid getting stuck in local suboptima and produce a more linear response to the control parameter $m$ (Fig.~\ref{figure:  uebersicht connectivity mit contour plots}).\\\\ 
The total costs expected to reach the same amount of conservation credits are significantly different between marginal and additive incentives in markets with low information, particularly in the transition regime (Fig.~\ref{figure: societal costs}). This strong cost increase originates from the too early abandonment of clusters under additive incentives. At increased levels of coordination, the cost differences between additive and marginal incentives are considerably reduced (Fig.~\ref{figure: societal costs coordination}). The most efficient allocation of conserved sites is achieved through collective decisions (cooperation), where the configuration space is explored in a way which avoids getting stuck in local minima (Fig.~\ref{figure: societal costs}) and differences between marginal and additive incentives disappear.

\subsection{Assumptions and generality of the results}
A prerequisite for the difference between marginal and additive incentives is that the ecosystem changes created by the market are not marginal. If changes were marginal, such that all interactions between sites are approximately unchanged by the actions induced by the market, one may simply use the marginal value as a basis for the evaluation, also in the case of fixed price payments schemes. For most real-world situations, however, it is likely that actions induced by the market interact. In such cases, it may be beneficial to implement incentives that explicitly include interactions between sites.\\\\
For such cases, we view the model presented in this paper as a generic model for the supply side of a spatial conservation market. The reader may have noted that we have fixed the supply, i.e. the amount of credits produced at each time step (eq.\ref{eq: Target Value}). One might suggest that the model should therefore be rather interpreted as the special case of a tradable permit market or an auction, where the amount of credits is fixed and prices vary. However, we could equally have fixed the costs of conservation, as was done with the same model in \citep{Hartig-Smartspatialincentives-2009}. Within the model assumptions, both options (fixing prices or quantities) are equivalent, because fixed prices result in a fixed amount of credits and vice versa. We therefore think that the model is generic for the way spatial incentives act on the spatial allocation of conservation measures and does not make any assumptions which would require that it be interpreted as a specific market form such as an auction, a payment scheme, or a permit market.\\\\ 
One assumption made in the model is that conservation costs are uncorrelated in time, as implied by the random sampling of costs from the uniform distribution at each step of the simulation process. While strong yearly changes of costs may occur in some situations, e.g. when farmers are using land that needs to rest for conservation, totally uncorrelated costs are, in general, rather unlikely. The results, however, qualitatively prevail in the case of cost correlations. Although not presented, we applied an unsystematic analysis of temporal and spatio-temporal correlation of costs as they were used in \citep{Hartig-Smartspatialincentives-2009}. Temporal correlation of costs acts similarly to increasing coordination steps in our model. This can be understood when considering that increasing coordination is essentially a temporal correlation of costs, as costs are fixed during the time of communication. Additional spatial cost correlation further decreases the efficiency losses associated with individual decisions. We assume that the reason for this is the spatial smoothing of the costs. When the cost correlation length is large compared to the correlation of the incentives, the chances of getting stuck in a local minimum decrease rapidly.\\\\
There are a number of abstractions from reality regarding the behavior of agents. First of all, for all three behavior models, we assumed that there are no additional costs associated with decisions and information exchange. Secondly, we assumed that agents are fair. In reality, both coordination through cheap talk and cooperation present possibilities to strategically exploit other agents. Cheap talk includes the possibility of strategic lies. Agents may try to induce their neighbors to make the first move towards conservation so that they themselves can subsequently free-ride. Cooperation requires landowners to reveal their true costs and temporarily accept lower payoffs if, in exchange, the group payoff is increased. Payments between landowners (side payments) could compensate for these losses, but the question of how this kind of cooperative system should be organized and whether it is stable against exploitation by defectors and free-riders remains open. One could, however, hope that stabilizing mechanisms such as reputation \citep{Fehr-theoryoffairness-1999, Sigmund-Rewardandpunishment-2001, Milinski-Reputationhelpssolve-2002} may improve cooperation at least when defection is observable for neighbors. However, if defection is only detectable indirectly, support for persistent cooperation would be much weaker. Experimental studies as proposed in \citet{Hartig-EcoTRADE-multi-2009} could help to study which of these strategies are most likely to be realized by human agents.\\\\
Finally, very important for real-world conservation are restoration costs and time lags for restoration. In this model, we assumed that land use can be changed instantaneously and without costs. The inclusion of costs and the time lags associated with habitat restoration appears to be an interesting problem for future research.   
\subsection{Policy implications}
Market-based instruments are increasingly applied for biodiversity conservation on private lands. At present, conservation markets seldom include spatial interactions between sites. One reason for this is that the inclusion of interactions increases transaction costs, i.e. costs associated with organizing the trading \citep{Salzman-Creatingmarketsecosystem-2005, Jack-Designingpaymentsecosystem-2008}. On the other hand, even simple spatial incentives may provide considerable efficiency gains for maintaining biodiversity by means of market-based instruments \citep{Hartig-Smartspatialincentives-2009}. Our results suggest that it is possible to control larger scale landscape structures with relatively simple spatial incentives. We tested two different spatial incentive mechanisms. Marginal incentives are dependent on the order of trading and therefore rather qualify for markets where trading takes place sequentially and the contract length is long, such as, for example, in tradable permit markets. Additive incentives are slightly simpler and qualify better for schemes where transactions are performed in parallel and the contract length is short, such as yearly payments. In the case of perfectly cooperating landowners, both options yield the same result. If landowners decide individually, however, marginal incentives generally perform better than additive incentives.\\\\
Therefore, the two main messages for policy are: 1) The reaction of landowners to spatial incentives is likely to differ with social organization and the applied incentive mechanism. 2) Besides changing the incentives, a market design that encourages cooperation seems beneficial when applying spatial metrics in market-based instruments.

\section{Acknowledgments}
The authors would like to thank Silvia Wissel, Karin Johst, Beatriz Vidal, Frank Wätzold, Astrid van Teeffelen, Volker Grimm, Horst Malchow, Heiko Rauhut and an anonymous reviewer for valuable comments and suggestions on the manuscript. The model description is inspired by the ODD protocol as proposed by \citet{Grimm-standardprotocoldescribing-2006}.

\appendix
\section{Optimimzation}\label{sec: Optimization}
To model a perfectly rational cooperating group of agents, we performed a global optimization of the land configuration. We used a slightly modified simulated annealing algorithm, which delivered better results than the original algorithm by \citet{Kirkpatrick-OptimizationBySimulated-1983}. The goal of the optimization was to minimize the total conservation costs as defined in eq.~\ref{def:societal costs} under the constraint of satisfying the fixed demand as given in eq.~\ref{eq: Target Value}:
\begin{equation}\label{eq: optimal allocation}
    \mathrm{min} \left\{ \sum_{i=1}^{n} \sigma_i \cdot c_i(t) \right \}_{U=\lambda \cdot N } \;.
\end{equation}
A random site $x_i$ was occupied with probability 
\begin{equation}\label{eq:simmulated annealing transition probabilities}
    p(\sigma_i,T)=\mathrm{min}\left(1,e^{- \frac{\phi_i(\sigma_i)-\Phi}{T}}\right)
\end{equation}
where $\phi_i(\sigma_i)$ is the marginal benefit-cost ratio of site $x_i$ and $\Phi$ is the average benefit-cost ratio of the present configuration. The simulated annealing was performed in $n$ steps with a new random subset of $N\cdot \nu$ of the sites at each step. To satisfy the constraints, each step was followed by adding (removing) conserved sites starting with the sites of highest (lowest) $\phi$ until the target value of $U=\lambda \cdot N $ was met. Temperature decay was exponential with decay parameter $\tau$ per time step. Table~\ref{table:optimization variables} lists the parameter values used.

\begin{table}[]
  \centering
  \begin{tabular}{lll} \toprule
  \textsc{Parameter} & \textsc{Connotation} & \textsc{Value}  \\ \midrule \addlinespace[0.2cm] 
  $T_0$ & Start temperature& 3  \\ 
  $\tau$ & Decay parameter & 0.001  \\ 
  $n$ & Steps & 3500   \\ 
  $\nu$ & Fraction of sites & 0.5  \\ \bottomrule

\end{tabular}
  \caption{Optimization parameters chosen for the simulated annealing}\label{table:optimization variables}
\end{table}

\end{document}